\documentclass[aps,prl,twocolumn,showpacs,superscriptaddress]{revtex4-2}  % for review and submission
\usepackage{graphicx}  % needed for figures
\usepackage{dcolumn}   % needed for some tables
\usepackage{bm}        % for math
\usepackage{amssymb}   % for math

\usepackage{xcolor}

\begin{document}
	
	\title{Precursor Film Spreading during Liquid Imbibition in Nanoporous Photonic Crystals}
	
	%\title{Spreading of Liquid precursor films on the Nanoscale:\\ Experiments with Nanoporous Photonic Band Gap Structures}
	
	\author{Luisa G. Cencha}
	\affiliation{Polymer Reaction Engineering Group, INTEC (Universidad Nacional del Litoral-CONICET), Güemes 3450, Santa Fe 3000, Argentina}
	
	\author{Guido Dittrich}
	\affiliation{Hamburg University of Technology, Materials Physics and High-Resolution X-Ray Analytics, Hamburg University of Technology, 21073 Hamburg, Germany}
	
	\author{Patrick Huber}
	\affiliation{Hamburg University of Technology, Materials Physics and High-Resolution X-Ray Analytics, Hamburg University of Technology, 21073 Hamburg, Germany}
	\affiliation{Deutsches Elektronen-Synchrotron DESY, Center for X-Ray and Nano Science, 22603 Hamburg, Germany}
	\affiliation{University of Hamburg, Centre for Hybrid Nanostructures CHyN, 22607 Hamburg, Germany}
	
	\author{Claudio L. A. Berli}
	\affiliation{INTEC (Universidad Nacional del Litoral-CONICET), Predio CCT CONICET Santa Fe, RN 168, 3000 Santa Fe, Argentina}
	
	\author{Raul Urteaga}
	\affiliation{IFIS-Litoral (Universidad Nacional del Litoral-CONICET), Guemes 3450, 3000 Santa Fe, Argentina}
	
	\date{\today}
	\begin{abstract}
		When a macroscopic droplet spreads, a thin precursor film of liquid moves ahead of the advancing liquid-solid-vapor contact line. Whereas this phenomenon has been explored extensively for planar solid substrates, its presence in nanostructured geometries has barely been studied so far, despite its importance for many natural and technological fluid transport processes. Here we use porous photonic crystals in silicon to resolve by light interferometry capillarity-driven spreading of liquid fronts in pores of few nanometers in radius. Upon spatiotemporal rescaling the fluid profiles collapse on master curves indicating that all imbibition fronts follow a square-root-of-time broadening dynamics. For the simple liquid (glycerol) a sharp front with a widening typical of Lucas-Washburn capillary-rise dynamics in a medium with pore-size distribution occurs. By contrast, for a polymer (PDMS) a precursor film moving ahead of the main menisci entirely alters the nature of the nanoscale transport, in agreement with predictions of computer simulations.
	\end{abstract}

	\maketitle
	
	The spreading of precursor films ahead of droplets of non-volatile fluids has been reported a century ago \cite{Hardy1919} and then widely studied in experiment, theory and simulations \cite{degennes1990spreading, Bonn2009}. Its dynamics follows a square-root-of-time scaling \cite{degennes1990spreading}, i.e. the radius $R$ of the wetted area by this precursor film increases with time $t$ as $R(t) \sim \sqrt{t}$, suggesting that the spreading process follows a simple diffusion process \cite{Abraham2002, Cazabat1997, Heslot1989a, Biance2004, MAlteraifi2018}. \textcolor{black}{The objective of this work is to investigate precursor film spreading during fluid imbibition in nanostrutured geometries. This is particularly important to understand the dynamics of the capillary filling of polymers in nanoporous media  \cite{Shin2007, Cuenca2013, Huber2015, Cao2016, Yao2018b, Cencha2018, Hor2018b}. Effectively,} when polymers move through pores having radii comparable to the polymer molecular size, a purely macroscopic analysis of the flow dynamics ignoring the impact of precursor film spreading can result to erroneous conclusions on confinement effects and predictions about the formed mesosocpic structures as well as characteristic processing times during imbibition \cite{Steinhart2002, Zhang2006, Martin2012, Huber2015}.
	
	However, the extreme challenges to monitor liquid fronts in nanoscale geometries have resulted in only a few experimental studies on imbibition fronts in nanoporous structures \cite{Huber2015, Gruener2012, Engel2010,Khaneft2013, Gruener2015, Vincent2016}. Engel and Stühn \cite{Engel2010} studied imbibition kinetics of polyisobutylene and poly-$\epsilon$-caprolactone in polycarbonate nanopores by time resolved small-angle X-ray scattering (SAXS). For both fluids a precursor film is inferred on short time scales, followed by a much longer complete filling of the pores by an advancing meniscus. Also using SAXS measurements Khaneft \textit{et al.} \cite{Khaneft2013} found that the imbibition of polystyrene in carbon nanotubes presents two different time scales which are related to the formation of a precursor film and the subsequent filling of the pore volume. The precursor formation was found to be six times faster than the volume filling, but no details about the precursor film profiles could be inferred. Qualitatively, these experimental findings confirmed hydrokinetic Lattice-Boltzmann (LB) and Molecular Dynamics (MD) simulations of the capillary filling in nano-channels by Chibbaro \textit{et al.} \cite{Chibbaro2008}. They clearly observed the formation of thin precursor films moving ahead of the main capillary front. The dynamics of the precursor films obeyed a $\sqrt{t}$ law as the main capillary front, although with a larger pre-factor than the main meniscus movement.
	\textcolor{black}{Moreover, additional possible scenarios exist where different types of thickness profiles can be found at the liquid-solid interface depending on the solid-liquid interaction \cite{Popescu2012}.}

	Here we report a light interferometry experiment that was specifically designed to elucidate the presence of precursor films during liquid imbibition into nanoporous media with mean pore sizes of 4 nm and 54$\%$ porosity \cite{SuppMat}. More precisely, we study capillary imbibition of a simple liquid (glycerol \cite{SuppMat}) and a complex, polymeric liquid (polydimethylsiloxane,  PDMS \cite{SuppMat}) in nanoporous silicon membranes. \textcolor{black}{  The viscosity $\mu$ and surface tension $\sigma$ of PDMS are  $5.2 Pa.s$ \cite{SuppMat} and $ 20.4 mN/m$\cite{Andriot}, respectively. While the respective values for glycerol are $0.953 Pa.s$ and $59.5 mN/m$ \cite{noauthororeditor2007handbook}}. The interferometric technique is based on the measurement of the membranes reflectance during liquid imbibition. The light reflectance spectra can be directly transformed to optical thickness and hence to the fluid distribution inside the pores. Thus, the measurement of the membranes' optical properties as a function of time allows the assessment of the fluid dynamics in the porous structure. 
	
\begin{figure}
		\includegraphics[scale=1]{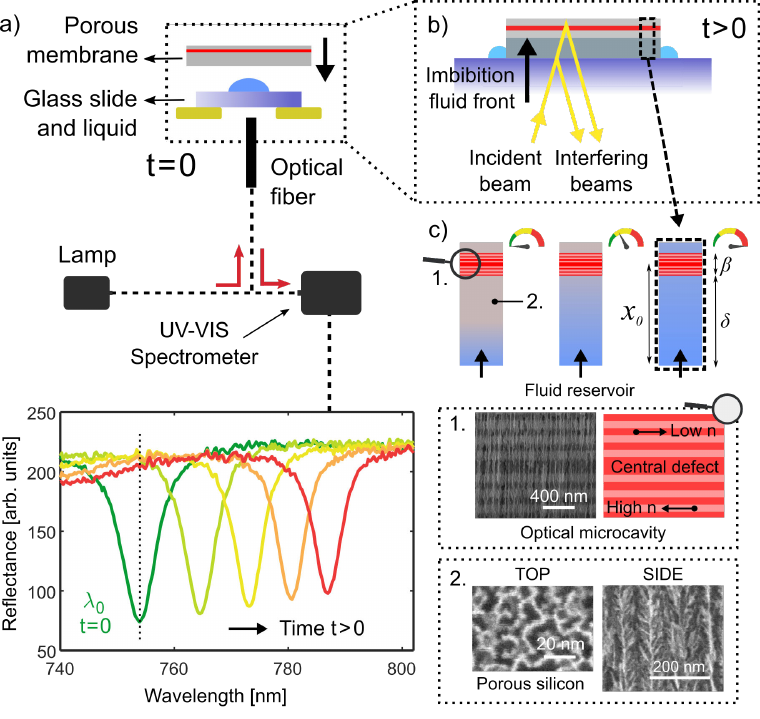}
		\caption{\label{fig:setup} Optofluidic imbibition experiments with nanoporous photonic crystals. (a) Experimental set-up. A UV-VIS spectrometer was used to acquire the light reflectance spectra as a function of time during imbibition. A liquid drop was placed onto a glass slide, and the porous membrane was hold on the top of the drop. When the acquisition of spectra began, the porous membrane was immediately placed onto the liquid drop (b).  (c) A porous silicon microcavity formed by alternating layers of two porosities (two refractive indexes) was used to detect the filling fraction inside the nanopores at the particular position of the central microcavity region.}
	\end{figure}

	A description of the interferometric technique can be found in Ref. \cite{Cencha2018}. The experimental set-up is sketched in Fig. \ref{fig:setup}a. The reflectance of the porous membrane is measured during imbibition using an UV-VIS-NIR spectrometer (Ocean Optics HR400) and a R400-7-SR fiber-optic reflection probe. The measurement of the optical thickness of homogeneous porous membranes has been used in previous works to quantify the overall amount of liquid infiltrating the structure \cite{Cencha2018,Cencha2019, Acquaroli2011,Urteaga2013}, without the ability to discriminate the spatial distribution inside the pores, i.e. the shape of the liquid front profile. 
	
	\textcolor{black}{In this work we propose the use of a photonic crystal to elucidate the spatiotemporal evolution of the wetting profile in pore space.}
	To that end, an optical microcavity was fabricated immediately after the homogeneous porous membrane ends. A description of the fabrication method can be found in Ref.  \cite{Acquaroli2011}. An optical microcavity is a one-dimensional photonic crystal composed of alternating layers with different refractive indexes and a central defect (Fig.~\ref{fig:setup}c). Due to the characteristic intensity amplification in the cavity at the resonance wavelength $\lambda_{\rm 0}$, the optical microcavity works as a sensor for liquid filling at the particular position of the microcavity center. The resonance wavelength $\lambda_{\rm 0}$ is solely determined by the optical thickness of the central defect. \textcolor{black}{ In turn, the optical thickness of this central porous layer depends on the liquid filling fraction, that is, on the relative proportion of liquid in the porous structure.} This means that during imbibition the resonance condition will shift to larger wavelengths only when the fluid reaches the microcavity's central defect, as shown in Fig.~\ref{fig:setup}a.
	\textcolor{black}{Using different initial porous layers (thicknesses $\delta=$ 3, 6, and 9 $\mu m$) to locate the center of the  microcavity (size $\beta=3\mu m$)  at different positions $x_{\rm{ 0}}=\delta +\beta/2$, allows one to monitor the filling fraction at a given position of the porous layer (see Fig. \ref{fig:setup}c).}
	
	\textcolor{black}{The optical properties of the membrane are closely related to the amount of fluid in the nanoporous matrix. The effective refractive index of the medium varies with the relative material composition (solid/air/liquid) that changes upon liquid imbibition.} By using effective medium theories \cite{Thei1997, Sallese2020}, the filling fraction in the microcavity can be obtained from the variation of the resonance position as, $ f(t)=\frac{\lambda(t)-\lambda_{\rm 0}}{\lambda_{\rm f}-\lambda_{\rm 0}}$, where $\lambda(t)$ is the resonance wavelength at time $t$, while $\lambda_{0}$ and $\lambda_{\rm f}$ are the resonance wavelengths at times $t=0$ and $t=\infty$, respectively. Thus, monitoring the resonance evolution enables a simple assessment of the filling fraction as a function of time in the microcavity central layer.
	
\begin{figure}
		\includegraphics[scale=0.85]{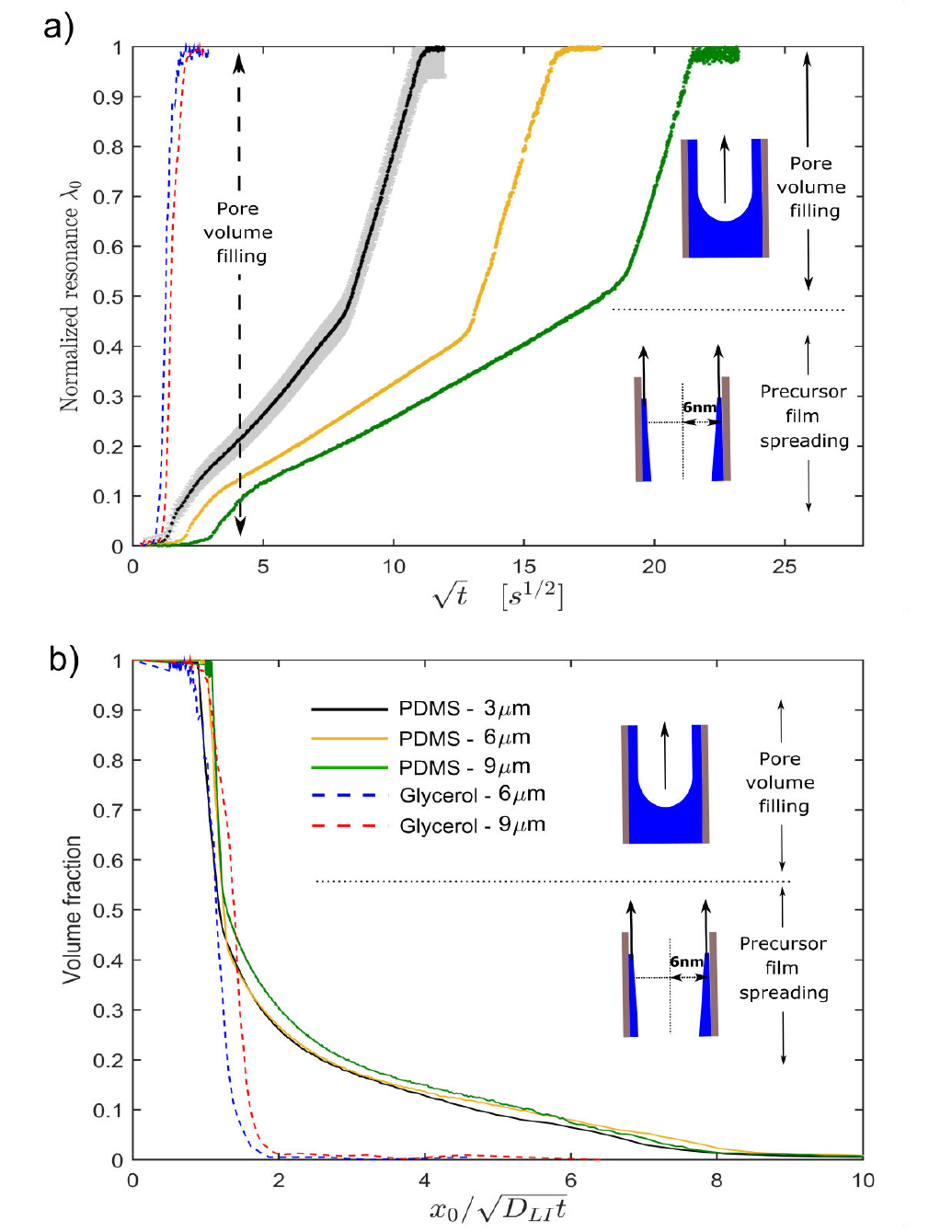}
		\caption{\label{fig:experiments} (a) Normalized resonance wavelength of the photonic crystal versus $\sqrt{t}$ for the imbibition of PDMS (dots) and glycerol (dashed lines) for different positions $x_{\rm 0}$ in the porous silicon membranes. \textcolor{black}{Error bars are indicated in the curve corresponding to PDMS in a porous layer of $\delta=3\mu m$. The insets schematically show the main flow regimes observed in experiments. (b) Pore filling fraction (proportional to the normalized $\lambda_{\rm 0}$) as a function of the dimensionless variable $x_{\rm 0}/\sqrt{D_{LI}t}$ for the imbibition of PDMS (solid lines) and glycerol (dashed lines). Note that, in the case of simple liquids, the complete filling of the pore volume fraction is expected at $x_{\rm 0}/\sqrt{D_{LI}t} = 1$ \cite{SuppMat}.}
		}
	\end{figure}

	The filling fraction of PDMS and glycerol obtained for different positions of the microcavity in the porous silicon structure are shown in Fig.~\ref{fig:experiments}a. The measured $f(t)$ is plotted versus $\sqrt{t}$ to better discriminate the distinct infiltration stages. 
	
	Figure \ref{fig:experiments}a indicates that: (1) The PDMS imbibition results in a front profile that maintains its shape,  \textcolor{black}{when different initial porous layer thickness $\delta$ are used. Moreover, three regimes can be discriminated: (\textit{i}) an initial stage below 5 $s^{1/2}$}, (\textit{ii}) a progressive filling during the mid stage, the onset of which is delayed according to the layers thickness, and (\textit{iii}) a more rapid filling during the last stage, the onset of which suggests the full covering of pore wall by the precursor film. Actually, the \textcolor{black}{ third} regime corresponds to the capillary filling of pre-wetted pores. (2) Only  \textcolor{black}{this third} regime can be seen for the case of glycerol, which indicates that pre-wetting does not take place, or it is below the resolution limit of our technique for this fluid, \textcolor{black}{and that potential pore-corners' effects \cite{Raeini2018} are not relevant in these experiments.} (3) The third regime for PDMS imbibition is composed of straight lines displaced by time intervals $\Delta t$, which are consistent with the Lucas-Washburn imbibition $\sqrt{t}$-dynamics followed by fluids in homogeneous porous membranes \cite{Cencha2018, Acquaroli2011,Urteaga2013}. \textcolor{black}{The initial shape (\textit{i}) of the curves for PDMS resembles the height profile of droplets of non-volatile liquids spreading on a smooth silicon substrate reported by Heslot \textit{et. al} \cite{Heslot1989a,Cazabat1990,Cazabat1997}. The rounded shape may correspond to the edge of the precursor film, which is limited by the minimum thickness attainable, as determined by the molecular size of the polymer.}
	
	These observations indicate also that the shape of the liquid front is always the same, but "stretched" by a $\sqrt{t}$-dependence. Thus, our experimental results support the simulation findings of Chibbaro \textit{et al.} \cite{Chibbaro2008}, where a diffusive behaviour was observed in capillary filling. Under this hypothesis, the shape of the advancing liquid front can be estimated from data in Fig.~\ref{fig:experiments}a. \textcolor{black}{To this end, Fig. \ref{fig:experiments}b presents the filling fraction as a function of the dimensionless variable $x_{\rm 0}/\sqrt{D_{LI}t}$, where  $D_{LI} = \langle r_h \rangle\sigma \cos{\theta}/(2\mu \tau^2)$ is the coefficient for liquid imbibition. In this expression $\langle r_h \rangle=6nm$ is the area wheighted mean hydrodynamic radius, $\theta\sim 0$ is the meniscus contact angle and $\tau=2.6$ is the tortuosity of the porous layer \cite{SuppMat}.}
	Indeed the spatially and temporal renormalized curves collapse on each other for each distinct fluid, corroborating the hypothesis that the same profile shape is present for all measured positions. \textcolor{black}{It is worth noting, however, that the viscosity value required to scale the PDMS curves was $44 Pa.s$. This value is significantly larger than the bulk viscosity reported above. Thus in agreement with previous observations on polymeric self-diffusion dynamics in nanopores, in particular in the pore-wall proximity \cite{Kusmin2010, Martin2010, Krutyeva2013}, confinement seems to reduce the fluidity of PDMS.}        
	
	%\begin{figure}
	%\includegraphics[scale=0.9]{Fig3.png}
	%\caption{\label{fig:mastercurve} Pore filling fraction (proportional %to the normalized $\lambda_{\rm 0}$) as a function of the reduced %variable $x_{\rm 0}/\sqrt{t}$ for the imbibition of PDMS (solid %lines) and glycerol (dashed lines) in the central layer of a %microcavity placed on the back of porous silicon membranes having %different thicknesses.}
	%\end{figure}
	
\begin{figure}
		\includegraphics[scale=0.85]{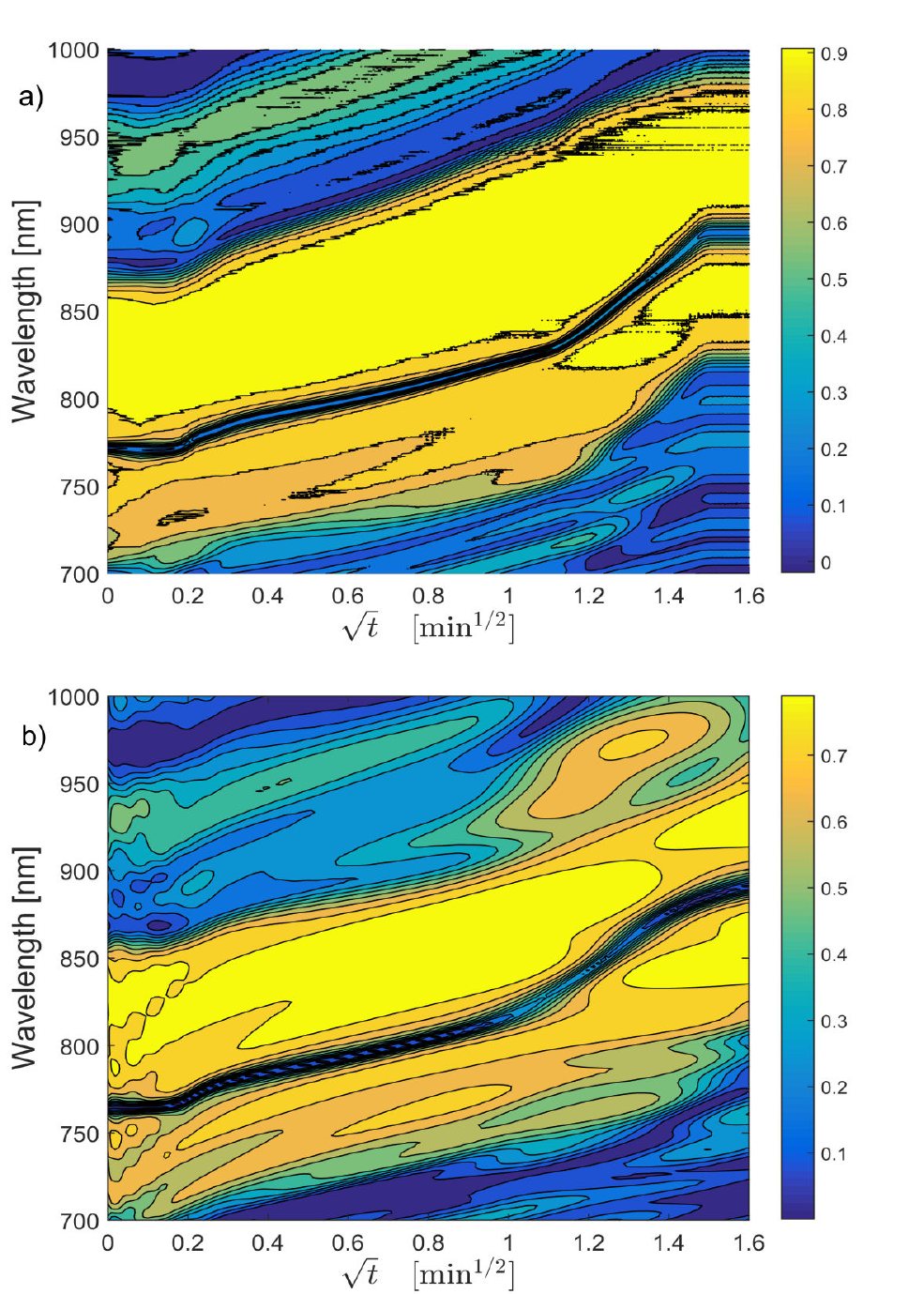}
		\caption{\label{fig:modelo} Experimental and simulated evolution of reflectance spectra upon imbibition of PDMS in a porous silicon layer having $\delta = 3\mu m$ thickness in front of a microcavity. a) Experimental reflectance evolution upon liquid imbibition. b) Simulated reflectance evolution upon liquid imbibition. The computations were made by using the master profile obtained in Fig.\ref{fig:experiments}b and parameters obtained by fitting the initial reflectance spectrum without any additional fitting parameters.}
	\end{figure}

	To additionally verify that the proposed experimental technique is adequate to predict the capillary filling dynamics within the porous structure, we simulated the reflectance spectrum that would be obtained using the filling profile shown in Fig. \ref{fig:experiments}b. The computation was performed using a matrix method \cite{Pedrotti}, while defining the optical properties of the layers using a simple effective medium model that accounts for the filling fraction at each position in the structure. The obtained results are shown in Fig. \ref{fig:modelo}. The simulated spectrum was obtained by fitting the porosities and thickness of the different layers of the porous substrate (see Fig. \ref{fig:setup}) to reproduce the reflectance spectrum before imbibition \cite{SuppMat}.  The fitting parameters were then used to simulate the complete evolution of the reflectance spectrum (\ref{fig:modelo}a and \ref{fig:modelo}b) using the master profile obtained in Fig. \ref{fig:experiments}b. A visual comparison corroborates the diffusion-like front shape evolution derived above in a rigorous manner. 
	
	It is interesting to note that a similar diffusion-like behavior as observed here in a nanoporous medium was found for spreading of a nonvolatile liquid (squalane) along wettable nanostripes embedded in a nonwettable surface using dynamic atomic force microscopy \cite{Checco2009}. In that case, variable spreading velocities were achieved using different amounts of liquid deposited in the reservoir pad, corroborating that the fluid front profile does not depend on the front speed, but only on the relative position in the channel.
	
	\textcolor{black}{Also in wetting experiments using PDMS ($\mu=1Pa\cdot s$) on flat silicon surfaces, Heslot et al. \cite{Feslot_1989}  found that the film spreading evolves with a diffusion-like coefficient $D \sim 3.7 \cdot 10^{-10}\frac{m^2}{s}$. To put our findings in the context of this work, we need to estimate the diffusion-like coefficient associated to the liquid film spreading, here defined as $D_{FS}$. As an approximation, $D_{FS}$ can be extracted from data in Fig. \ref{fig:experiments}b; more precisely, from the region that corresponds to the advance of the precursor film alone. The tangent to the curve in that region intercepts the abscissa axis at $x_{\rm 0}/\sqrt{D_{LI}t}\sim 8$. Therefore, one obtains $D_{FS}= 8^2 \cdot D_{LI}\sim 1.3\cdot 10^{-11}m^2/s$. 
		To compare this value with that obtained on flat substrates, the tortuosity of our porous silicon structure must be considered; then, $\tau^2 D_{FS}\sim 9\cdot 10^{-11}m^2/s$. Despite the fact that neither the experimental setup nor the characteristics of the working PDMS were the same, this coefficient is quite similar to the one found in Ref. \cite{Feslot_1989}.}
	Hence, our experiments in a nanoporous solid are not only in a qualitative manner, but also in a quantitative manner in good agreement with spreading dynamics of PDMS in planar silicon geometry. This encourages us to believe that other effects than precursor film spreading on the single-channel scale that could result in a $\sqrt{t}$ spreading of the imbibition front are not dominating the kinetics here. In particular, the pore-size distribution along with the Lucas-Washburn $\sqrt{t}$ of menisci movements in non-interconnected single-pores could also result in a $\sqrt{t}$ broadening, where the prefactor (the diffusion coefficient) would be determined by the width of the pore-size distribution \cite{Gruener2012, Gruener2015,Sallese2020}. However, since we employed the identical nanoporous geometries with identical pore-size distributions this broadening should occur for both fluids in an identical manner, which is obviously not the case.
	
	In summary, we could demonstrate that porous photonic band gap structures are particularly suitable for studying liquid imbibition at the nanoscale. For glycerol, a fluid with comparably small molecules, we find a sharp, step-like fluid profile. By contrast the imbibition with the macromolecular PDMS is significantly affected by precursor film spreading, in agreement with  simulation studies \cite{Chibbaro2008}, experiments at planar surfaces \cite{MAlteraifi2018} and in tubular nanopores \cite{Engel2010, Khaneft2013} for polymeric liquids. %\textcolor{black}{In addition, our observation of the squared root dependence could be another indication of reduced entanglement densities in nanopore-confined polymeric systems, in accordance with previous imbibition \cite{Shin2007} and quasi-elastic neutron scattering experiments \cite{Martin2010, Krutyeva2013}}.
	
	For the future more detailed MD simulations, also as a function of the complexity of the imbibing liquid and nanoporous geometries could be  helpful to achieve an improved understanding of the interplay of imbibition and imbibition front broadening in nanoporous media. Moreover, we envision that the optofluidic technique employed here could be also of interest to study liquid distributions during drying of porous media, where a complex rearrangement encompassing thin-layer flow at pore walls and flow in the pore centre inside, couples to evaporative transport outside of the porous medium \cite{Chauvet2009,Shokri2019}.
	
	Finally, liquid-infused, most prominently polymer-filled nanoporous solids play an increasingly important role in technologies ranging from drug delivery to the design of functional materials for energy storage and harvesting \cite{Steinhart2002, Huber2015, Brinker2020}. Our study substantially improves the possibilites to rationally design these systems by combining nanofluidics with modern optics.    
	
	\nocite{Looyenga1965,Rabbani2014} 
	
	\begin{acknowledgements}
		Funding by the Deutsche Forschungsgemeinschaft (DFG), Nos. 192346071 (SFB 986, TP B7), 422879465 (SPP 2171) and 383411810 is gratefully acknowledged.
	\end{acknowledgements}

	\bibliography{Manuscript}
	
\end{document}